\documentclass[11pt,twoside,onecolumn]{article}
\usepackage{amsfonts}
\usepackage{eurosym}
\usepackage{latexsym}
\usepackage{epsfig}
\usepackage{amsmath,amssymb}
\usepackage[dvips]{color}

\newcommand{\be}{\begin{eqnarray}}
\newcommand{\ee}{\end{eqnarray}}

\begin{document}

\title{\textbf{Brane-world stars with solid crust and vacuum exterior}}
\author{Jorge Ovalle$^{1}$\thanks{%
jovalle@usb.ve}, $\,$ L\'{a}szl\'{o} \'{A}. Gergely$^{2,3,4}$\thanks{%
laszlo.a.gergely@gmail.com} $\,$ and Roberto~Casadio$^{5,6}$\thanks{%
casadio@bo.infn.it} \\
\null
\\
$^{1}$\emph{Departamento de F\'{\i}sica, Universidad Sim\'{o}n Bol\'{\i}var} 
\\
\emph{Apartado 89000, Caracas 1080A, Venezuela} \\
$^{2}$\emph{Department of Theoretical Physics, University of Szeged} \\
\emph{Tisza Lajos krt 84-86, Szeged 6720, Hungary}\\
$^{3}$\emph{Department of Experimental Physics, University of Szeged,} \\
\emph{D\'{o}m T\'{e}r 9, Szeged 6720, Hungary}\\
$^{4}$\emph{Department of Physics, Tokyo University of Science, 1-3} \\
\emph{Kagurazaka, Shinjyuku-ku, Tokyo 162-8601, Japan} \\
$^{5}$\emph{Dipartimento di Fisica e Astronomia, Universit\`{a} di Bologna,} 
\\
\emph{via Irnerio~46, 40126 Bologna, Italy }\\
$^{6}$\emph{Istituto Nazionale di Fisica Nucleare, Sezione di Bologna,} \\
\emph{via B.~Pichat~6/2, 40127 Bologna, Italy} }
\maketitle

\begin{abstract}
The minimal geometric deformation approach is employed to show the existence
of brane-world stellar distributions with vacuum Schwarzschild exterior,
thus without energy leaking from the exterior of the brane-world star into
the extra dimension. The interior satisfies all elementary criteria of
physical acceptability for a stellar solution, namely, it is regular at the
origin, the pressure and density are positive and decrease monotonically
with increasing radius, finally all energy conditions are fulfilled. A very
thin solid crust with negative radial pressure separates the interior from
the exterior, having a thickness $\Delta $ inversely proportional to both
the brane tension $\sigma $ and the radius $R$ of the star, i.e. $\Delta
^{-1}\sim R\,\sigma $. This brane-world star with Schwarzschild exterior
would appear only thermally radiating to a distant observer and be fully
compatible with the stringent constraints imposed on stellar parameters by
observations of gravitational lensing, orbital evolutions or properties of
accretion disks.
\end{abstract}

%
%
%
%
%
%
%
%
%
%
%
%

\section{Introduction}

\setcounter{equation}{0} 

The Randall-Sundrum (RS) second brane-world model~\cite{RS} is based on
perceiving our 4-dimensional space-time as a hypersurface embedded into the
5-dimensional bulk. The observed world is induced on it by virtue of a
discontinuity in the extrinsic curvature (expressed by the
Lanczos-Sen-Darmois-Israel junction conditions~\cite{Lanczos}-\cite{Israel}).
Despite efforts spent in recent years for understanding the inner
workings of the model, the possible impact of the fifth dimension on the
4-dimensional gravity sector is still not fully assessed.

The original RS model was generalised to allow for a codimension one curved
brane embedded into a generic 5-dimensional bulk (for a general review see
Ref.~\cite{MaartensLivRev}). At the level of the formalism the gravitational
dynamics was worked out in its full generality, either in the covariant $4+1$
(brane plus extra-dimension) decomposition~\cite{SMS}-\cite{Decomp2}, in the
$3+1+1$ canonical~\cite{s+1+1}-\cite{s+1+1Ham} or in the $3+1+1$
covariant~\cite{Maartens}-\cite{3+1+1} approaches.
 
A necessary ingredient in such theories is the brane tension. Its huge value
has been constrained from below by employing tabletop measurements of the
gravitational constant~\cite{tabletop}-\cite{EotWash}, imposing the condition
that during a cosmological evolution at the time of Nucleosynthesis brane
effects must already be severely dampened~\cite{nucleosynthesis}, or from
astrophysical considerations~\cite{germ}. 

Further cosmological investigations revealed important modifications in the
early universe~\cite{BDEL} as compared to standard cosmology. The thermal
radiation of an initially very hot brane could even lead to black hole
formation in the fifth dimension~\cite{ChKN}. Such a black hole modifies the
5-dimensional Weyl curvature, backreacting onto the curvature and the
dynamics of the brane. Structure formation was analysed in
Refs.~\cite{PalStructure}-\cite{PalStructure2}. Unfortunately the lack of closure of
the dynamical equations on the brane, despite some
progress~\cite{perturb}-\cite{perturb3}, hindered the development of a full
perturbation theory on the brane, hence observations on the Cosmic Microwave
Background and on structure formation could not be explored in full generality
within the theory. Confrontation with Nucleosynthesis~\cite{Nucleosynthesis} and
type~Ia supernova data has been successfully done~\cite{supernova}-\cite{supernova2}.
Brane-world effects were also shown to successfully replace dark matter,
both in the dynamics of clusters~\cite{HarkoClusters} and in galactic
dynamics~\cite{HarkoRC}-\cite{BoehmerHarko}, thus solving the rotation
curve problem~\cite{rotcurv}.

Although many fundamental aspects in the RS scenario, in particular the
cosmological aspects, were clarified (largely in the sense to push the
high-energy modifications so close to the Big Bang that their effects remain
unobservable), certain key issues remain unresolved, which are mostly
related to self-gravitating systems and black holes, and for which the
high-energy regimes could be within observational reach. 

The simplest, spherically symmetric brane solution is similar to the general
relativistic Reissner-Nordstr\"{o}m solution, but the role of the square of
the electric charge is taken by a tidal charge originating in the higher
dimensional Weyl curvature~\cite{dadhich}. The value of the tidal charge was
constrained by confronting with observations in the
Solar System~\cite{BHL}-\cite{TidalLens}.
A rotating generalisation is also known~\cite{Aliev}. Gravitational collapse
on the brane has been investigated in Refs.~\cite{BGM}-\cite{gergely2007}.
Early work on stellar solutions and astrophysics in brane-world context can
be found in Refs.~\cite{germ}, \cite{ND}-\cite{Gregory2006}. The topic is
reviewed in Ref.~\cite{MaartensLivRev}.

There is some evidence indicating the existence of RS black hole
metrics~\cite{FW11}-\cite{LBH13}, but an exact solution of the full set of dynamical
equations of the 5-dimensional gravity has not been discovered so far.
Solving the full 5-dimensional Einstein field equations has indeed proven an
extremely complicated task (see, e.g.~\cite{cmazza}-\cite{darocha2012}, and
references therein). Beside, such a black hole solution could exhibit
various facets in our 4-dimensional world, as there are many possible ways
to embed a 4-dimensional brane into the 5-dimensional bulk to get a
4-dimensional section of it. While it is true that the $\mathbb{Z}_{2}$
symmetry across the brane is employed as a common simplifying assumption, it
is not mandatory either, and lifting it leads to further freedom in the
embedding, explored in Refs.~\cite{lamasym1}-\cite{perk} (different
cosmological constants), \cite{bhasym}-\cite{DVDP} (different 5-dimensional
black hole masses) or~\cite{both1}-\cite{BCG} (both).

The study of exact, physically acceptable solutions to the effective
4-dimensional Einstein field equations on the brane could clarify certain
aspects of the 5-dimensional geometry and provide hints on the ways our
observed universe could be embedded into it. When starting from any brane solution,
the Campbell-Magaard theorems~\cite{campbell}-\cite{sss} can be employed to
extend them, at least locally, into the bulk. Consequences of the
Campbell-Magaard theorems for General Relativity (GR) were discussed in
Refs.~\cite{maia95}-\cite{dahia03}. The rigorous study of the effective Einstein
field equations on the 4-dimensional brane therefore qualifies as a first
step of this process. It also helps clarifying the role of the 5-dimensional
contributions to the sources of the effective 4-dimensional field equations.

Deriving physically acceptable exact solutions in GR is
an extremely difficult task, even in vacuum~\cite{Stephani}, due to the
complexity of the Einstein field equations. For inner stellar solutions, the
task is even more complicated (a useful algorithm to obtain all static
spherically symmetric perfect fluid solutions in GR was presented
in~\cite{lake03}, and its extension to locally anisotropic fluids
in~\cite{herrera08}). Only a small number of internal solutions are
known~\cite{lake98}. This complexity is further amplified in brane-worlds,
where nonlinear terms in the matter fields appear as high-energy corrections.

A useful guide is provided by the requirement that GR should be recovered at
low energies, where it has been extensively tested. Fortunately, as we mentioned
above, the RS theory contains a free parameter, the brane tension $\sigma$,
which allows to control this important aspect by precisely setting the scale of high
energy physics~\cite{jovalle07}. This fundamental requirement stands at the
basis of the \textit{minimal geometric deformation approach\/}
(MGD)~\cite{jovalle2009}, which has made possible, among other things,
to derive exact and physically acceptable solutions for spherically
symmetric~\cite{jovalle207} and non-uniform stellar distributions~\cite{jovalleprd13},
to generate other physically acceptable inner stellar
solutions~\cite{jovalle07}-\cite{jovalleBWstars}, to express the tidal charge in
the metric found in Ref.~\cite{dadhich} in terms of the Arnowitt-Deser-Misner
mass, to study microscopic black holes~\cite{covalle1}-\cite{covalle2},
to elucidate the role of exterior Weyl stresses (arising from bulk gravitons)
acting on compact stellar distributions~\cite{olps2013}, as well as extend the concept
of variable tension introduced in~\cite{Decomp2}, \cite{Eotvos} by analyzing
the behaviour~\cite{cor2014} of the black strings extending into the extra
dimension~\cite{gergely2006}.

For spherically symmetric systems on the brane, the RS scenario provides two
quantities of extra-dimensional origin, namely, the dark radiation
$\mathcal{U}$ and dark pressure $\mathcal{P}$, which act as sources of
4-dimensional gravity even in vacuum. How the various choices of these
quantities affect stellar structures on the brane is only partially understood
so far~\cite{kanti2013}-\cite{garcia2014}. Most remarkably, the Schwarzschild
exterior can be generated by a static self-gravitating star only if it consists of a
certain exotic fluid. By contrast, if the stellar system contains regular
matter, there must be an exchange of energy between its 4-dimensional
exterior with the 5-dimensional bulk, leading to non-static configurations.
Since the Schwarzschild geometry is not the exterior of brane-world stars
consisting of regular matter, the modifications in the exterior geometry due
to extra-dimensional effects have been one of the main targets of
investigations in the search for ways to find evidence of extra-dimensional
gravity.

Nevertheless, the Schwarzschild geometry is strongly supported by weak-field
tests of gravity in the Solar System, constraining such possible
extra-dimensional modifications (for a recent work on classical test of GR
in the brane-world context see Ref.~\cite{cuzzi2014}).
According to Ref.~\cite{Tanaka}, black holes in the RS brane-world should evaporate
by Hawking radiation, thus the existence of long-lived solar mass black holes could
constrain the bulk curvature radius.
The same conclusion was reached in Ref.~\cite{EFK}, so it seems that the price
to pay for a static exterior would be Hawking radiation, which has however a temperature
smaller than the temperature of the CMB for objects of a few solar masses.

In this paper, we shall employ the MGD approach to show that, despite severe
constraints on brane-world stars from available data, there are scenarios in
which their existence cannot be ruled out. We shall show that certain
brane-world star exteriors might be the same as in GR, hence automatically
fitting all observational constraints. In particular, with the exception of
a tiny solid crust, the stellar matter could have the most reasonable
physical properties and matching conditions are obeyed between the interior
geometry and the exterior Schwarzschild vacuum.

The paper is organised as follows. In Section~\ref{II}, we briefly review
the effective Einstein field equations on the brane for a spherically
symmetric and static distribution of matter with density $\rho $, radial
pressure $p_{r}$ and tangential pressure $p_{t}$. We also present the MGD
approach. In Section~\ref{III}, we show that it is possible to have a star
made of regular matter with a Schwarzschild exterior and without exchange of
energy between the brane and the bulk at the price of including a tiny solid
stellar crust. Finally, we summarise our conclusions.

\section{Effective Einstein equations and MGD}

\setcounter{equation}{0} \label{II} In the generalised RS brane-world
scenario, gravitation acts in five dimensions and modifies gravitational
dynamics in the (3+1)-dimensional
universe accessible to all other physical fields, the so called brane. The
arising modified Einstein equations (with $G$ the 4-dimensional Newton
constant, $k^{2}=8\,\pi \,G$, and $\Lambda $ the 4-dimensional cosmological
constant) 
\begin{equation}
G_{\mu \nu }=-k^{2}\,T_{\mu \nu }^{\mathrm{eff}}-\Lambda \,g_{\mu \nu }\ ,
\label{4Dein}
\end{equation}%
could formally be seen as Einstein equations in which the energy-momentum
tensor $T_{\mu \nu }$ is complemented by new source terms, which contribute
to an effective source as 
\begin{equation}
T_{\mu \nu }\ \rightarrow \ T_{\mu \nu }^{\mathrm{eff}}=T_{\mu \nu }+\frac{6%
}{\sigma }\,S_{\mu \nu }+\frac{1}{8\pi }\,\mathcal{E}_{\mu \nu }+\frac{4}{%
\sigma }\mathcal{F}_{\mu \nu }\ .  \label{tot}
\end{equation}%
Here $\sigma $ is again the brane tension, 
\begin{equation}
S_{\mu \nu }=\frac{T\,T_{\mu \nu }}{12}-\frac{T_{\mu \alpha }\,T_{\ \nu
}^{\alpha }}{4}+\frac{g_{\mu \nu }}{24}\left( 3\,T_{\alpha \beta
}\,T^{\alpha \beta }-T^{2}\right)
\end{equation}%
represents a high-energy correction quadratic in the energy-momentum tensor (%
$T=T_{\alpha }^{\ \alpha }$), 
\begin{equation}
k^{2}\,\mathcal{E}_{\mu \nu }=\frac{6}{\sigma }\left[ \mathcal{U}\left(
u_{\mu }\,u_{\nu }+\frac{1}{3}\,h_{\mu \nu }\right) +\mathcal{P}_{\mu \nu }+%
\mathcal{Q}_{(\mu }\,u_{\nu )}\right]
\end{equation}%
is a non-local source, arising from the 5-dimensional Weyl curvature (with $%
\mathcal{U}$ the bulk Weyl scalar; $\mathcal{P}_{\mu \nu }$ and $\mathcal{Q}%
_{\mu }$ the stress tensor and energy flux, respectively), and $\mathcal{F}%
_{\mu \nu }$ contains contributions from all non-standard model fields
possibly living in the bulk (it does not include the 5-dimensional
cosmological constant, which is fine-tuned to $\sigma $ in order to generate
a small 4-dimensional cosmological constant). For simplicity, we shall
assume $\mathcal{F}_{\mu \nu }=0$ and $\Lambda =0$ throughout the paper.

Let us then restrict to spherical symmetry (such that $\mathcal{P}_{\mu \nu
}=\mathcal{P}\,h_{\mu \nu }$ and $\mathcal{Q}_{\mu }=0$) and choose as the
source term in Eq.~(\ref{tot}) a perfect fluid, 
\begin{equation}
T_{\mu \nu }=(\rho +p)\,u_{\mu }\,u_{\nu }-p\,g_{\mu \nu }\ ,
\label{perfect}
\end{equation}%
where $u^{\mu }=e^{-\nu /2}\,\delta _{0}^{\mu }$ is the fluid 4-velocity
field in the Schwarzschild-like coordinates of the metric 
\begin{equation}
ds^{2}=e^{\nu (r)}\,dt^{2}-e^{\lambda (r)}\,dr^{2}-r^{2}\left( d\theta
^{2}+\sin ^{2}\theta \,d\phi ^{2}\right) \ .  \label{metric}
\end{equation}%
Here $\nu =\nu (r)$ and $\lambda =\lambda (r)$ are functions of the areal
radius $r$ only, ranging from $r=0$ (the star's centre) to some $r=R$ (the
star's surface).

The metric~(\ref{metric}) must satisfy the effective 4-dimensional Einstein
field equations~\eqref{4Dein}, which here read~\cite{covalle2} 
\begin{eqnarray}
&&k^{2}\left[ \rho +\strut \displaystyle\frac{1}{\sigma }\left( \frac{\rho
^{2}}{2}+\frac{6}{k^{4}}\,\mathcal{U}\right) \right] =\strut \displaystyle%
\frac{1}{r^{2}}-e^{-\lambda }\left( \frac{1}{r^{2}}-\frac{\lambda ^{\prime }%
}{r}\right)  \label{ec1} \\
&&  \notag \\
&&k^{2}\strut \displaystyle\left[ p+\frac{1}{\sigma }\left( \frac{\rho ^{2}}{%
2}+\rho \,p+\frac{2}{k^{4}}\,\mathcal{U}\right) +\frac{4}{k^{4}}\frac{%
\mathcal{P}}{\sigma }\right] =-\frac{1}{r^{2}}+e^{-\lambda }\left( \frac{1}{%
r^{2}}+\frac{\nu ^{\prime }}{r}\right)  \label{ec2} \\
&&  \notag \\
&&k^{2}\strut \displaystyle\left[ p+\frac{1}{\sigma }\left( \frac{\rho ^{2}}{%
2}+\rho \,p+\frac{2}{k^{4}}\mathcal{U}\right) -\frac{2}{k^{4}}\frac{\mathcal{%
P}}{\sigma }\right] =\frac{1}{4}e^{-\lambda }\left[ 2\,\nu ^{\prime \prime
}+\nu ^{\prime 2}-\lambda ^{\prime }\,\nu ^{\prime }+2\,\frac{\nu ^{\prime
}-\lambda ^{\prime }}{r}\right] \ ,  \label{ec3}
\end{eqnarray}%
with primes denoting derivatives with respect to $r$. Moreover, 
\begin{equation}
p^{\prime }=-\strut \displaystyle\frac{\nu ^{\prime }}{2}(\rho +p)\, ,
\label{con1}
\end{equation}%
The 4-dimensional GR equations are recovered for $\sigma ^{-1}\rightarrow 0$%
, and the conservation equation~(\ref{con1}) then becomes a linear
combination of Eqs.~(\ref{ec1})-(\ref{ec3}).

By simple inspection of the field equations (\ref{ec1})-(\ref{ec3}), we
identify the effective density $\tilde{\rho}$, effective radial pressure $%
\tilde{p}_{r} $ and effective tangential pressure $\tilde{p}_{t}$, which are
given by 
\begin{equation}
\tilde{\rho}\,=\,\rho +\strut \displaystyle\frac{1}{\sigma }\left( \frac{%
\rho ^{2}}{2}+\frac{6}{k^{4}}\,\mathcal{U}\right) \,,  \label{efecden}
\end{equation}%
\begin{equation}
\tilde{p}_{r}\,=\,p+\frac{1}{\sigma }\left( \frac{\rho ^{2}}{2}+\rho \,p+%
\frac{2}{k^{4}}\,\mathcal{U}\right) +\frac{4}{k^{4}}\frac{\mathcal{P}}{%
\sigma }\,,  \label{efecprera}
\end{equation}%
\begin{equation}
\tilde{p}_{t}\,=\,p+\frac{1}{\sigma }\left( \frac{\rho ^{2}}{2}+\rho \,p+%
\frac{2}{k^{4}}\,\mathcal{U}\right) -\frac{2}{k^{4}}\frac{\mathcal{P}}{%
\sigma }\,,  \label{efecpretan}
\end{equation}%
clearly illustrating that extra-dimensional effects produce anisotropies in
the stellar distribution, that is 
\begin{equation}
\Pi \equiv \tilde{p}_{r}-\tilde{p}_{t} = \frac{6}{k^{4}}\frac{\mathcal{P}}{%
\sigma } \ .
\end{equation}%
A GR isotropic stellar distribution (perfect fluid) therefore becomes an
anysotropic stellar system on the brane.

Next, the MGD approach~\cite{jovalle2009} will be introduced in order to
generalise GR interior solutions to the brane-world scenario.

\subsection{Star interior from MGD}

Eqs.~(\ref{ec1})-(\ref{con1}) represent an open system of differential
equations on the brane. For a unique solution additional information on the
bulk geometry and on the embedding of the 4-dimensional brane into the bulk
is required~\cite{FW11}-\cite{cmazza}, \cite{darocha2012}.

In its absence, one can rely on the MGD induced by a GR solution. In order
to implement the MGD approach, we first rewrite the field equations~%
\eqref{ec1}-\eqref{ec3} as 
\begin{eqnarray}
&&e^{-\lambda }=1-\frac{k^{2}}{r}\int_{0}^{r}x^{2}\left[ \rho +\frac{1}{%
\sigma }\left( \frac{\rho ^{2}}{2}+\frac{6}{k^{4}}\,\mathcal{U}\right) %
\right] dx  \label{lam} \\
&&  \notag \\
&&\frac{1}{k^{2}}\,\frac{\mathcal{P}}{\sigma }=\frac{1}{6}\left( G_{\
1}^{1}-G_{\ 2}^{2}\right)  \label{pp} \\
&&  \notag \\
&&\frac{6}{k^{4}}\,\frac{\mathcal{U}}{\sigma }=-\frac{3}{\sigma }\left( 
\frac{\rho ^{2}}{2}+\rho \,p\right) +\frac{1}{k^{2}}\left( 2\,G_{\
2}^{2}+G_{\ 1}^{1}\right) -3\,p\ ,  \label{uu}
\end{eqnarray}%
where 
\begin{equation}
G_{\ 1}^{1}=-\frac{1}{r^{2}}+e^{-\lambda }\left( \frac{1}{r^{2}}+\frac{\nu
^{\prime }}{r}\right)  \label{g11a}
\end{equation}%
and 
\begin{equation}
G_{\ 2}^{2}=\frac{1}{4}\,e^{-\lambda }\left( 2\,\nu ^{\prime \prime }+\nu
^{\prime 2}-\lambda ^{\prime }\,\nu ^{\prime }+2\frac{\nu ^{\prime }-\lambda
^{\prime }}{r}\right) \ .  \label{g22}
\end{equation}

Now, by using Eq.~\eqref{uu} in Eq.~\eqref{lam}, we find an
integro-differential equation for the function $\lambda (r)=\lambda (\nu
(r),r)$, which is different from the GR case, and is a direct consequence of
the non-locality of the brane-world equations. The general solution to this
equation is given by~\cite{jovalle2009} 
\begin{eqnarray}
e^{-\lambda } &\!\!=\!\!&\underbrace{{1-\frac{k^{2}}{r}\int_{0}^{r}x^{2}\,%
\rho \,dx}}_{\mathrm{GR-solution}}+\underbrace{e^{-I}\int_{0}^{r}\frac{e^{I}%
}{\frac{\nu ^{\prime }}{2}+\frac{2}{x}}\left[ H(p,\rho ,\nu )+\frac{k^{2}}{%
\sigma }\left( \rho ^{2}+3\,\rho \,p\right) \right] dx+\beta (\sigma
)\,e^{-I},}_{\mathrm{Geometric\ deformation}}  \notag \\
&\!\!\equiv \!\!&\mu (r)+f(r)\ ,  \label{edlrwss}
\end{eqnarray}%
where 
\begin{equation}
H(p,\rho ,\nu )\equiv 3\,k^{2}\,p-\left[ \mu ^{\prime }\left( \frac{\nu
^{\prime }}{2}+\frac{1}{r}\right) +\mu \left( \nu ^{\prime \prime }+\frac{%
\nu ^{\prime 2}}{2}+\frac{2\nu ^{\prime }}{r}+\frac{1}{r^{2}}\right) -\frac{1%
}{r^{2}}\right]  \label{finalsol}
\end{equation}%
encodes the effects due to bulk gravity, depending on $p$, $\rho $ and $\nu $%
. The exponent 
\begin{equation}
I\equiv \int \frac{\left( \nu ^{\prime \prime }+\frac{{\nu ^{\prime }}^{2}}{2%
}+\frac{2\nu ^{\prime }}{r}+\frac{2}{r^{2}}\right) }{\left( \frac{\nu
^{\prime }}{2}+\frac{2}{r}\right) }\,dr\ ,  \label{I}
\end{equation}%
and $\beta =\beta (\sigma )$ is a function of the brane tension $\sigma $
which must vanish in the GR limit. Moreover, in the star interior, the
condition $\beta =0$ must be imposed in order to avoid singular solutions at
the center $r=0$. Finally, note that the function 
\begin{equation}
\mu (r)\equiv 1-\frac{k^{2}}{r}\int_{0}^{r}x^{2}\,\rho \,dx=1-\frac{2\,m(r)}{%
r}  \label{standardGR}
\end{equation}%
contains the usual GR mass function $m$.

An important remark is that when a given (spherically symmetric) perfect
fluid solution in GR is considered as a candidate solution for the
brane-world system of Eqs.~\eqref{ec1}-\eqref{con1} \lbrack or,
equivalently, Eq.~\eqref{con1} along with Eqs.~\eqref{lam}-\eqref{uu}], one
exactly obtains 
\begin{equation}
H(p,\rho ,\nu )=0\ .  \label{H=0}
\end{equation}%
Therefore, every (spherically symmetric) perfect fluid solution in GR will
produce a \textit{minimal\/} deformation on the radial metric component~(\ref%
{edlrwss}), such that the geometric deformation $f=f(r)$ contains only one
contribution, and 
\begin{equation}
f^{\ast }(r)=\frac{2\,k^{2}}{\sigma }\,e^{-I(r)}\int_{0}^{r}\frac{x\,e^{I(x)}%
}{x\,\nu ^{\prime }+4}\left( \rho ^{2}+3\,\rho \,p\right) dx\ .
\label{fsolutionmin}
\end{equation}%
The geometric deformation $f=f(r)$ of Eq.~\eqref{edlrwss} ``distorts'' the
GR solution represented by Eq.~\eqref{standardGR}, but the specific form
$f^{\ast}=f^{\ast }(r)$ in Eq.~\eqref{fsolutionmin} represents a ``minimal
distortion'' for any GR solution of choice in the sense that all of the
deforming terms in Eq.~(\ref{edlrwss}) have been removed, except for
those produced by the density and pressure, which will always be present
in a realistic stellar distribution~\footnote{There is a MGD solution in the
case of a dust cloud, with $p=0$, but we will
not consider it in the present work.}.
Note then that the function $f^{\ast }$ can also be
found from Eq.~(\ref{pp}) as 
\begin{equation}
\frac{6}{k^{2}}\frac{\mathcal{P}}{\sigma }=\left( \frac{1}{r^{2}}+\frac{\nu
^{\prime }}{r}-\frac{\nu ^{\prime \prime }}{2}-\frac{{\nu ^{\prime }}^{2}}{4}%
-\frac{\nu ^{\prime }}{2\,r}\right) f^{\ast }-\frac{1}{4}\bigg(\nu ^{\prime
}+\frac{2}{r}\bigg){(f^{\ast })}^{\prime }\ .  \label{ppf3}
\end{equation}

The interior stellar geometry is given by the MGD metric 
\begin{equation}
ds^{2}=e^{\nu ^{-}(r)}\,dt^{2}-\frac{dr^{2}}{1-\frac{2\,m(r)}{r}+f^{\ast }(r)%
}-r^{2}\left( d\theta ^{2}+\sin {}^{2}\theta d\phi ^{2}\right) \ ,
\label{mgdmetric}
\end{equation}%
and it is straightforward to introduce the effective interior mass function 
\begin{equation}
\tilde{m}(r)=m(r)-\frac{r}{2}\,f^{\ast }(r) \ .  \label{effecmass}
\end{equation}%
Since, from Eq.~(\ref{fsolutionmin}), the geometric deformation in Eq.~(\ref%
{mgdmetric}) is seen to obey the positivity condition 
\begin{equation}
f^{\ast }(r)\geq 0 \ ,  \label{f*>0}
\end{equation}
the effective interior mass~(\ref{effecmass}) is always reduced by the
extra-dimensional effects.

\subsection{Interior MGD metric and exterior Weyl fluid}

The MGD metric in Eq.~(\ref{mgdmetric}), characterising the star interior at 
$r<R$, should be matched with an exterior geometry associated with the Weyl
fluid $\mathcal{U}^{+}$, $\mathcal{P}^{+}$, and $p=\rho =0$, for $r>R$~\cite%
{germ}. This can be generically written as 
\begin{equation}
ds^{2}=e^{\nu ^{+}(r)}\,dt^{2}-e^{\lambda ^{+}(r)}\,dr^{2}-r^{2}\left(
d\theta ^{2}+\sin {}^{2}\theta d\phi ^{2}\right) \ ,  \label{genericext}
\end{equation}%
where the explicit form of the functions $\nu ^{+}$ and $\lambda ^{+}$ are
obtained by solving the effective 4-dimensional vacuum Einstein equations,
namely 
\begin{equation}
R_{\mu \nu }-\frac{1}{2}\,g_{\mu \nu }\,R_{\ \alpha }^{\alpha }=\mathcal{E}%
_{\mu \nu }\qquad \Rightarrow \qquad R_{\ \alpha }^{\alpha }=0\ ,
\end{equation}%
where we recall that extra-dimensional effects are contained in the
projected Weyl tensor $\mathcal{E}_{\mu \nu }$ and that only a few
analytical solutions are known to date~\cite{dadhich}, \cite{CFMsolution}, 
\cite{FW11}-\cite{cmazza}. When both interior and exterior metrics,
respectively given by Eq.~(\ref{mgdmetric}) and (\ref{genericext}) are
considered, continuity of the first fundamental form at the star surface $%
\Sigma $ defined by $r=R$ reads~\cite{Israel}, \cite{israel} 
\begin{equation}
\left[ ds^{2}\right] _{\Sigma }=0\ ,  \label{match1}
\end{equation}%
where $[F]_{\Sigma }\equiv F(r\rightarrow R^{+})-F(r\rightarrow R^{-})\equiv
F_{R}^{+}-F_{R}^{-}$, for any function $F=F(r)$, and yields 
\begin{eqnarray}
e^{\nu ^{-}(R)} &\!\!=\!\!&e^{\nu ^{+}(R)}\ ,  \label{ffgeneric1} \\
&&  \notag \\
1-\frac{2\,M}{R}+f_{R}^{\ast } &\!\!=\!\!&e^{-\lambda ^{+}(R)}\ ,
\label{ffgeneric2}
\end{eqnarray}%
where $M=m(R)$. Likewise, continuity of the second fundamental form at the
star surface reads~\cite{Israel}, \cite{israel} 
\begin{equation}
\left[ G_{\mu \nu }\,r^{\nu }\right] _{\Sigma }=0\ ,  \label{matching1}
\end{equation}%
where $r_{\mu }$ is a unit radial vector. Using Eq.~\eqref{matching1} and
the general Einstein field equations, we then find 
\begin{equation}
\left[ T_{\mu \nu }^{eff}\,r^{\nu }\right] _{\Sigma }=0\ ,  \label{matching2}
\end{equation}%
which leads to 
\begin{equation}
\left[ p+\frac{1}{\sigma }\left( \frac{\rho ^{2}}{2}+\rho \,p+\frac{2}{k^{4}}%
\,\mathcal{U}\right) +\frac{4}{k^{4}}\,\frac{\mathcal{P}}{\sigma }\right]
_{\Sigma }=0\ .  \label{matching3}
\end{equation}%
Since we assume the distribution is surrounded by a Weyl fluid characterised
by $\mathcal{U}^{+}$, $\mathcal{P}^{+}$, and $p=\rho =0$ for $r>R$, this
matching condition takes the final form 
\begin{equation}
p_{R}+\frac{1}{\sigma }\left( \frac{\rho _{R}^{2}}{2}+\rho _{R}\,p_{R}+\frac{%
2}{k^{4}}\,\mathcal{U}_{R}^{-}\right) +\frac{4}{k^{4}}\frac{\mathcal{P}%
_{R}^{-}}{\sigma }=\frac{2}{k^{4}}\frac{\mathcal{U}_{R}^{+}}{\sigma }+\frac{4%
}{k^{4}}\frac{\mathcal{P}_{R}^{+}}{\sigma }\ ,  \label{matchingf}
\end{equation}%
where $p_{R}\equiv p_{R}^{-}$ and $\rho _{R}\equiv \rho _{R}^{-}$. Finally,
by using Eqs.~(\ref{uu}) and (\ref{ppf3}) in the condition~(\ref{matchingf}%
), the second fundamental form can be written in terms of the MGD at the
star surface, denoted by $f_{R}^{\ast }$, as 
\begin{equation}
p_{R}+\frac{f_{R}^{\ast }}{8\pi }\left( \frac{\nu _{R}^{\prime }}{R}+\frac{1%
}{R^{2}}\right) =\frac{2}{k^{4}}\frac{\mathcal{U}_{R}^{+}}{\sigma }+\frac{4}{%
k^{4}}\frac{\mathcal{P}_{R}^{+}}{\sigma }\ ,  \label{sfgeneric}
\end{equation}%
where $\nu _{R}^{\prime }\equiv \partial _{r}\nu _{-}|_{r=R}$. The
expressions given by Eqs.~(\ref{ffgeneric1}) and (\ref{ffgeneric2}), along
with Eq.~(\ref{sfgeneric}), are the necessary and sufficient conditions for
the matching of the interior MGD metric~(\ref{mgdmetric}) to a spherically
symmetric \textquotedblleft vacuum\textquotedblright\ filled by a
brane-world Weyl fluid.

The matching condition~(\ref{sfgeneric}) yields an important result: in the
Schwarzschild exterior one must have $\mathcal{U}^{+}=\mathcal{P}^{+}=0$,
which then leads to 
\begin{equation}
p_{R}=-\frac{f_{R}^{\ast }}{8\pi }\left( \frac{\nu _{R}^{\prime }}{R}+\frac{1%
}{R^{2}}\right) \ .  \label{pnegative}
\end{equation}%
Since we showed above, in Eq.~\eqref{f*>0}, that $f^{\ast }\geq 0$, an
exterior vacuum can only be supported in the brane-world by exotic stellar
matter, with negative pressure $p_{R}$ at the star surface, in agreement
with the model discussed in Ref.~\cite{gergely2007}.

\section{Stellar solution with a solid crust}

\setcounter{equation}{0} \label{III} In this section we will show that,
contrary to what was previously believed, it is possible to have a star
predominantly made of regular matter and with a Schwarzschild exterior.

\begin{figure}[t]
\centering
\epsfxsize=10cm \epsfbox{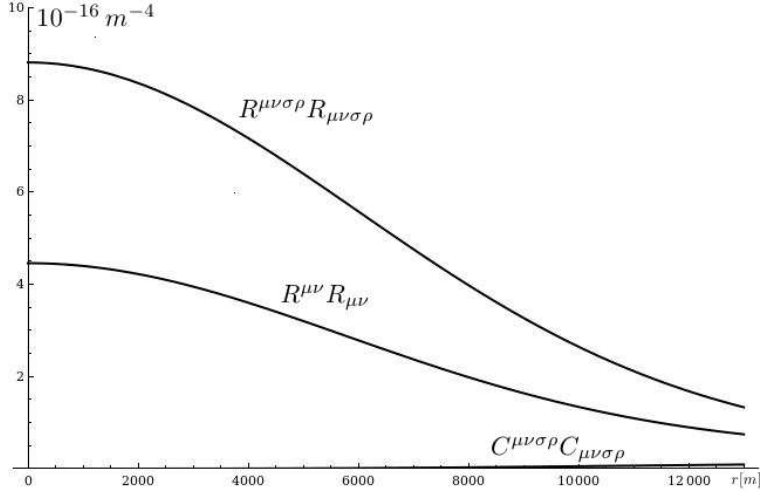}
\caption{The Kretschmann scalar $R^{\protect\mu \protect\nu \protect\sigma 
\protect\rho }R_{\protect\mu \protect\nu \protect\sigma \protect\rho }$,
Ricci square $R^{\protect\mu \protect\nu }R_{\protect\mu \protect\nu }$ and
Weyl square $C^{\protect\mu \protect\nu \protect\sigma \protect\rho }C_{%
\protect\mu \protect\nu \protect\sigma \protect\rho }$ for a distribution
with $R=13\,$km. }
\label{K1}
\end{figure}

In order to accomplish the above, we consider the exact interior brane-world
solution found in Ref.~\cite{jovalle207}, 
\begin{equation}
e^{\nu }=A(1+x)^{4}\ ,  \label{g00}
\end{equation}%
\begin{equation}
e^{-\lambda }=1-\frac{8}{7}\,x\frac{(3+x)}{(1+x)^{2}}+f^{\ast }(r)\ ,
\label{g11}
\end{equation}%
with $x = C\,r^2$ and matter pressure and density given by 
\begin{equation}
p(r)=\frac{2C(2-7x-x^2)}{7\pi (1+x)^{3}}\ ,  \label{regularpress}
\end{equation}%
\begin{equation}
\rho (r)=\frac{C\,\left( 9+2\,x+x^2\right) }{7\,\pi \,{\left( 1+x\right) }%
^{3}}\ ,  \label{regularden}
\end{equation}%
and non-local contributions

\begin{eqnarray}
\mathcal{P}(r) &=&\frac{32C^{2}}{441x^{2}(1+x)^{6}(1+3x)^{2}}\left[ x\left(
180+2040x+8696x^{2}\right. +16533x^{3}+12660x^{4}\right.   \notag
\label{regP} \\
&&\left. \left. +146x^{5}-120x^{6}+9x^{7}\right) -60\sqrt{C}%
(1+x)^{3}(3+26x+63x^{2})\mathrm{arctan}(\sqrt{x})\right] ,
\end{eqnarray}%
\begin{eqnarray}
\mathcal{U}(r) &=&\frac{32C^{2}}{441x^{2}(1+x)^{6}(1+3x)^{2}}\left[
x^{2}\left( 795+4865x+10044x^{2}+6186x^{3}\right. \right.   \notag
\label{regU} \\
&&\left. \left. -373x^{4}-219x^{5}-18x^{6}\right) -240x^{3/2}(1+x)^{3}(5+9x)%
\mathrm{arctan}(\sqrt{x})\right] .
\end{eqnarray}%
The corresponding geometric deformation in Eq.~(\ref{g11}) due to
5-dimensional effects is now given by 
\begin{equation}
f^{\ast }=\left( \frac{2}{7}\right) ^{2}\frac{C}{\sigma \,\pi }\left[ \frac{%
240+589x-25x^{2}-41x^{3}-3x^{4}}{3(1+x)^{4}(1+3x)}-\frac{80\,\mathrm{arctan}(%
\sqrt{x})}{(1+x)^{2}(1+3x)\sqrt{x}}\right] \ ,  \label{mgd}
\end{equation}%
with $A$ and $C$ constants to be determined by the matching conditions.
Using this interior brane-world solution, and the Schwarzschild exterior
geometry with $\mathcal{M}=\tilde{m}(R)$, 
\begin{equation}
e^{\nu ^{+}}=e^{-\lambda ^{+}}=1-\frac{2\mathcal{M}}{r}\ ,\qquad \mathcal{U}%
^{+}=\mathcal{P}^{+}=0\ ,  \label{SchwExt}
\end{equation}%
in the matching conditions~(\ref{ffgeneric1}) and (\ref{ffgeneric2}), we
obtain 
\begin{equation}
A=\left( 1-\frac{2\mathcal{M}}{R}\right) (1+CR^{2})^{-4}
\label{RegmatchSchw1}
\end{equation}%
and 
\begin{eqnarray}
\frac{2\mathcal{M}}{R} &\!\!=\!\!&\frac{2M}{R}-\left( \frac{2}{7}\right) ^{2}%
\frac{C}{\pi \,\sigma }\,\frac{%
240+589CR^{2}-25C^{2}R^{4}-41C^{3}R^{6}-3C^{4}R^{8}}{%
3(1+CR^{2})^{4}(1+3CR^{2})}  \notag \\
&&+\left( \frac{2}{7}\right) ^{2}\frac{C}{\pi \,\sigma }\,\frac{80\,\mathrm{%
arctan}(\sqrt{C}R)}{(1+CR^{2})^{2}(1+3CR^{2})\sqrt{C}R}\ .
\label{RegmatchSchw2}
\end{eqnarray}%
Using Eqs.~(\ref{g00}), (\ref{regularpress}) and (\ref{mgd}) in the matching
condition~(\ref{pnegative}), the constant $C$ turns out to be determined by 
\begin{equation}
CR^{2}\left( 2-7CR^{2}-C^{2}R^{4}\right) +\frac{7}{16}\left( 1+CR^{2}\right)
^{2}\left( 1+9CR^{2}\right) \,f_{R}^{\ast }(\sigma )=0\ ,
\label{RegmatchSchw3}
\end{equation}%
which clearly shows that $C$ is promoted to a function of the brane tension $%
\sigma $ due to bulk gravity effects, that is $C=C(\sigma )$. The
Kretschmann scalar $R^{\mu \nu \sigma \rho }R_{\mu \nu \sigma \rho }$, Ricci
square $R^{\mu \nu }R_{\mu \nu }$ and Weyl square $C^{\mu \nu \sigma \rho
}C_{\mu \nu \sigma \rho }$ associated with the interior geometry given by
the expressions~(\ref{g00}) and~(\ref{g11}) are shown on Fig.~\ref{K1}.

In order to find the extra-dimensional effects on physical variables, i.e.,
the pressure in Eq.~(\ref{regularpress}) and density in
Eq.~(\ref{regularden}), we need to fix the function $C=C(\sigma)$ satisfying
Eq.~(\ref{RegmatchSchw3}). We first consider 
\begin{equation}  \label{RegC}
C = C_0+\delta \ ,
\end{equation}
where $C_0$ is the GR value of $C$, given by 
\begin{equation}  \label{RegC0}
C_0 = \frac{\sqrt{57}-7}{2R^2} \ ,
\end{equation}
and which is found by using the standard GR condition at the star surface, $%
p(R)\equiv p_R=0$, in Eq.~(\ref{regularpress}). By using Eq.~(\ref{RegC}) in
Eq.~(\ref{RegmatchSchw3}), we then obtain the leading-order brane-world
contribution 
\begin{equation}  \label{deltaf}
\delta(\sigma) = \frac{7(1+C_0R^2)^2(1+9C_0R^2)}{16C_0R^2(7+2C_0R^2)} \frac{%
f^{*}_R}{R^2} + \mathcal{O}(\sigma^{-2}) \ .
\end{equation}
The pressure can thus be determined by expanding $p=p(C)$ around $C_0$, 
\begin{equation}  \label{ExpanP}
p(C_0+\delta)\, {\simeq} \, p(C_0)+\left.\delta\,{\frac{dp}{dC}}%
\right|_{C=C_0} \ ,
\end{equation}
which leads to 
\begin{equation}  \label{RegPSchw}
p(r) \, {\simeq} \, \frac{2C_0}{7\pi} \frac{(2-7C_0r^2-C_0^2r^4)}{%
(1+C_0r^2)^3} +\frac{4}{7\pi}\frac{(1-9C_0r^2+2C_0^2r^4)}{(1+C_0r^2)^4}
\delta(\sigma) \ .
\end{equation}
Consequently, at the star surface $r=R$, the pressure becomes 
\begin{equation}  \label{RegPSchw2}
p_R(\sigma) {\simeq} \frac{4}{7\pi}\frac{(1-9C_0R^2+2C_0^2R^4)}{(1+C_0R^2)^4}%
\, \delta(\sigma) < 0 \ ,
\end{equation}
that is, negative and proportional to $1/\sigma$, according to Eqs.~(\ref%
{mgd}) and (\ref{deltaf}).
\par 
\begin{figure}[t!]
\centering
\epsfxsize=10cm \epsfbox{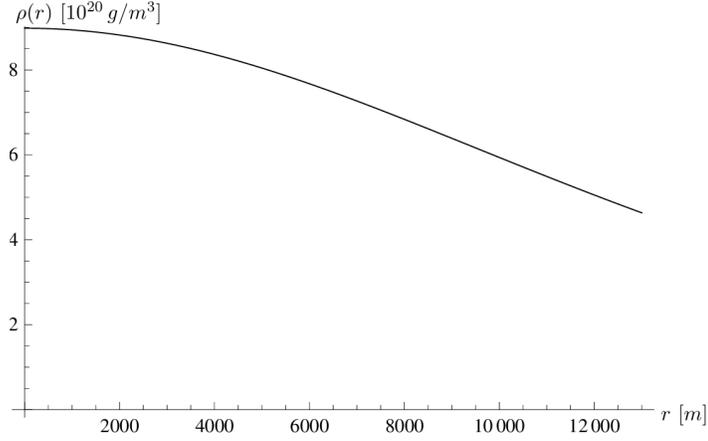}
\caption{Behaviour of the density $\protect\rho=\protect\rho(r)$ [in units
of $10^{20}\,$g/m$^3$] for $\protect\delta/ C_0 = 0.03$, for a typical
compact distribution of $R =13\,$km.}
\label{rho2}
\end{figure}
\begin{figure}[h!]
\centering
\epsfxsize=10cm \epsfbox{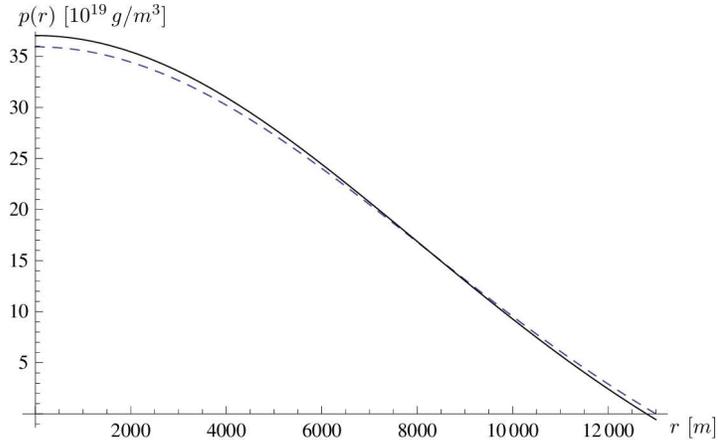}
\caption{Qualitative comparison of the pressure $p=p(r)$ [in units of
$10^{19}$\,g/m$^3$] for a compact distribution of $R =13\,$km in GR [$p(R)=0$;
dashed curve] and in the brane-world model with $\protect\delta/C_0 = 0.03$
[$p(R)<0$; solid curve] both with Schwarzschild exterior, showing the thin
layer of solid outer crust in the brane-world case.}
\label{p}
\end{figure}
A typical density profile $\rho=\rho(r)$ is displayed in Fig.~\ref{rho2}. In
Fig.~\ref{p}, we likewise display the pressure $p=p(r)$ for different values
of the radius, both for the GR case and for its brane-world generalisation.
Remarkably, the pressure is negative only in a thin layer close to the
boundary. A negative pressure in this layer acts as a positive tension, a
common property for solid materials. Hence we can interpret the structure of
the star as an (effectively) imperfect fluid with a solid crust. In this
respect, it is now important to look at the energy conditions for our system.
\par
Let us recall the energy conditions are a set of constraints which are
usually imposed on the energy-momentum tensor in order to avoid exotic
matter sources and, correspondingly, exotic space-time geometries. In
particular, we may mention: (a) the Null Energy Condition (NEC), $%
T_{\mu\nu}K^{\mu}K^{\nu}\geq 0$ for any null vector $K^{\mu}$. For a perfect
fluid, this condition implies 
\begin{eqnarray}
\rho+p\geq 0 \ ;  \label{NEC}
\end{eqnarray}
(b) the Weak Energy Condition (WEC), $T_{\mu\nu}X^{\mu}X^{\nu}\geq 0$ for
any time-like vector $X^{\mu}$, which, again for a perfect fluid, yields $%
\rho\geq 0$ and $\rho+p\geq 0$; (c) the Dominant Energy Condition (DEC), $%
T^{\mu}_{\ \nu}\,X^{\nu}=-Y^\mu$, where $Y^\mu$ must be a future-pointing
causal vector. For a perfect fluid, this means 
\begin{eqnarray}
\rho>\mid p\mid \ ;  \label{DEC}
\end{eqnarray}
and finally (d) the Strong Energy Condition (SEC), $(T_{\mu\nu}-\frac{1}{2}%
\,T\,g_{\mu\nu})X^\mu\,X^\nu\geq 0$, or, for a perfect fluid, $\rho+p\geq 0$
and $\rho+3\,p\geq 0$. While it is true that these conditions might fail for
particular reasonable classical systems, they can be viewed as sensible
guides to avoid unphysical solutions. A very well known example is the usual
classical fields which obey the WEC, and therefore the energy density seen
by any (time-like) observer can never be negative. Hence wormholes,
superluminal travel, and construction of time machines can be ruled out. On
the other hand, the SEC is violated by Cosmological Inflation (driven by a
minimally coupled massive scalar) and by the accelerating universe~\cite%
{Visser97}. Let us also note that DEC $\Rightarrow$ WEC $\Rightarrow$ NEC
and SEC $\Rightarrow$ NEC (but SEC does not imply WEC). We can now argue how
to implement these conditions in our case, where a GR isotropic fluid has
been transformed into an anysotropic one due to extra-dimensional effects,
as it is clearly seen from Eqs.~(\ref{efecden})-(\ref{efecpretan}). To
address this question, we shall consider the weakest condition, namely the
NEC, and show with a direct calculation the difference with respect to the
perfect fluid case. First of all, we need a null vector field $K^{\mu}$,
which in our case can be written as 
\begin{equation}
K^\mu = e^{-\nu/2}\,\delta_{0}^{\ \mu}+e^{-\lambda/2}\,\delta_{1}^{\ \mu} \ ,
\end{equation}
for which the NEC reads 
\begin{equation}
T_{\mu\nu}\,K^{\mu}\,K^{\nu} = e^\nu\,\tilde{\rho}\,K^0\,K^0 +e^\lambda\,%
\tilde{p}_r\,K^1\,K^1 = \tilde{\rho}+\tilde{p}_r \geq 0 \ ,
\end{equation}
which looks like the standard condition~\eqref{NEC} with $\rho\to\tilde{\rho}
$ and $p_r\to\tilde{p}_r$. In the same way, the DEC leads to $\tilde{\rho}%
\geq\tilde{p}_r$ and $\tilde{\rho}\geq\tilde{p}_t$, which are precisely the
analogue forms of Eq.~\eqref{DEC}.
\begin{figure}[t!]
\centering
\epsfxsize=10cm \epsfbox{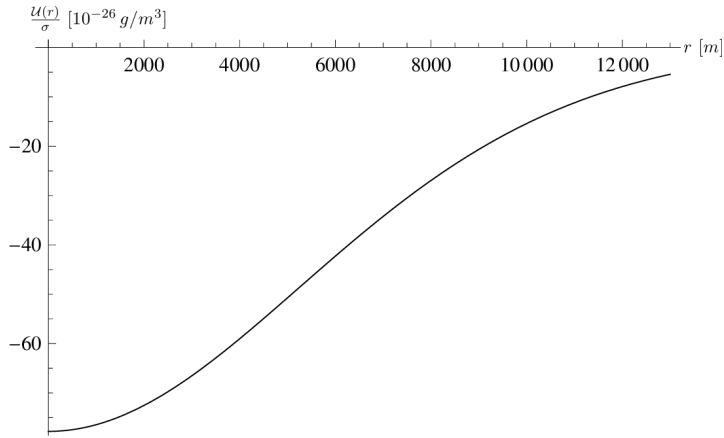}
\caption{The scalar function ${\mathcal{U}(r)}/{\protect\sigma}$ [in units
of $10^{-26}\,$g/m$^3$] for a distribution with $R=13\,$km. $\mathcal{U}(r)$
is always negative in the interior, hence it reduces both the effective
density and effective pressure.}
\label{anisoU}
\end{figure}
\begin{figure}[h!]
\centering
\epsfxsize=10cm \epsfbox{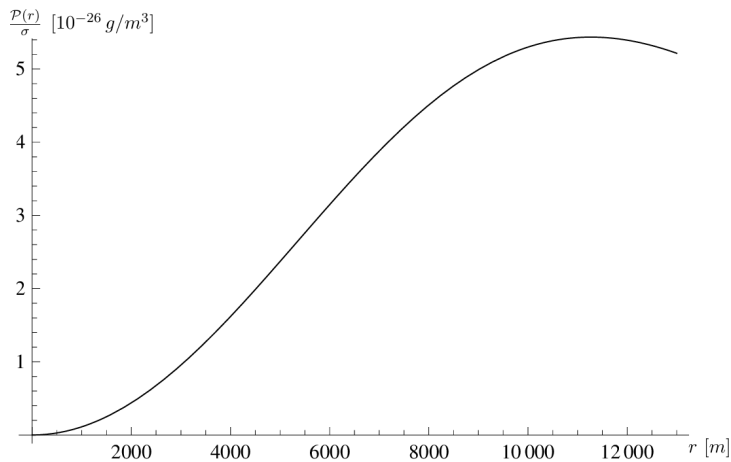}
\caption{Behaviour of the Weyl function ${\mathcal{P}(r)}/{\protect\sigma}$
[in units of $10^{-26}\,$g/m$^3$] inside the stellar distribution with $%
R=13\,$km. }
\label{anisoP}
\end{figure}

According to Eqs.~(\ref{efecden})-(\ref{efecpretan}), these inequalities
turn into new bounds for the prefect fluid density and pressure, 
\begin{equation}
\rho\, \geq\, p+\frac{1}{\sigma }\rho \,p +\frac{4}{k^{4}\sigma} (\mathcal{P}
-\mathcal{U}) \ ,
\end{equation}
\begin{equation}
\rho\, \geq\, p+\frac{1}{\sigma }\rho \,p -\frac{2}{k^{4}\sigma} (\mathcal{P}%
+2\,\mathcal{U}) \ ,
\end{equation}
while the effective strong energy condition becomes 
\begin{equation}
\rho+3\,p +\frac{1}{\sigma }\left( 2\,\rho^{2}+3\,\rho \,p+\frac{12}{k^{4}}\,%
\mathcal{U}\right) > 0 \ .
\end{equation}
All of the above effective conditions are satisfied, as well as the WEC,
even inside the solid crust. This means that there are no
negative (fluid or effective) pressures comparable in magnitude or larger
than the density $\rho$, and therefore the brane-world effects on the GR
solution are not strong enough to jeopardise the physical acceptability of
the system. In Figs.~\ref{anisoU} and~\ref{anisoP}, we also display the
behaviour of $\mathcal{U}=\mathcal{U}(r)$ and $\mathcal{P}=\mathcal{P}(r)$,
respectively, for the same star as in Fig.~\ref{p}. These two plots clearly
show the typical energy scale of the Weyl functions and a discontinuity at
$r=R$ in the respective quantities. We shall have more to say about this in
the last Section.

In order to determine the thickness $\Delta $ of the solid crust, we define
the critical radius $r_{c}$ as the areal radius of the sphere on which the
pressure vanishes (see Fig.~\ref{p} for an example), 
\begin{equation}
p(r_{c})=0\ .
\end{equation}%
Therefore, the crust has a thickness 
\begin{equation}
\Delta \equiv R-r_{c}\ ,  \label{D}
\end{equation}%
where $r_{c}$ can now be found by using the pressure in Eq.~(\ref%
{regularpress}), 
\begin{equation}
p(r_{c})=\frac{2\,C\,(2-7\,C\,r_{c}^{2}-C^{2}\,r_{c}^{4})}{7\pi
(1+C\,r_{c}^{2})^{3}}=0\ .  \label{crit1}
\end{equation}%
The above immediately yields 
\begin{equation}
r_{c}=\sqrt{\frac{\sqrt{57}-7}{2\,C(\sigma )}}\ ,  \label{crit2}
\end{equation}%
where $C(\sigma )$ is given by Eq.~(\ref{RegC}). To leading order in $\sigma
^{-1}$ this gives 
\begin{equation}
r_{c}\simeq R\left( 1+\frac{\delta }{C_{0}}\right) ^{-1/2}\simeq R\left( 1-%
\frac{\delta (\sigma )}{2\,C_{0}}\right) \ ,  \label{crit3}
\end{equation}%
and 
\begin{equation}
\Delta \simeq \frac{R\,\delta (\sigma )}{2\,C_{0}}\ ,
\end{equation}%
which, according to Eqs.~(\ref{RegC0}), (\ref{deltaf}) and~(\ref{mgd}),
finally reads 
\begin{equation}  \label{jgr}
\Delta \sim R^{3}\,\delta (\sigma )\sim R\,f_{R}^{\ast } \sim \frac{1}{%
R\,\sigma } \ .
\end{equation}

We emphasise that the tiny region with $p<0$ is interpreted a solid crust,
consisting of regular matter. A test particle at the star surface $r=R$
would experience a combination of a negative pressure $p(R)<0$ and
gravitational force, both pulling it inwards, and an extra-dimensional
effect pushing it out. In this sense, the negative pressure of the crust
resembles a fluid tension in a soap bubble. One can consider our solution in
analogy with the structure of neutron stars, which have a solid crust
surrounding a (superfluid) interior.

The expression in Eq.~(\ref{jgr}) shows that the solid crust becomes thicker as
the size of the star becomes smaller, showing that this ``solidification
process'' in the outer layer due to extra-dimensional effects should be
particularly important for compact distributions. Note however that for
solar size stars $R\gg \sigma^{-1/2}$, and the crust is much thinner than
the fundamental length $\sigma^{-1/2}$. This already suggests that the crust
is of little physical relevance, if not a pure artefact of the
approximations employed, as we shall discuss shortly.

\section{Conclusions}

\setcounter{equation}{0} 

Our main result in this paper is that a brane-world compact source, despite
previous no-go results, may have a Schwarzschild exterior. The exact
solution of the effective Einstein equations on the brane derived here,
represents a non-uniform, spherically symmetric, self-gravitating star with
regular properties, which is embedded into a vacuum Schwarzschild exterior
geometry. This result is obtained at the price of having negative pressure
inside a narrow shell at the star surface [$p(r)<0$ for $R-\Delta <r<R$].
Throughout the star (in the shell and interior), however, all physical
properties are perfectly regular.
\begin{figure}[t!]
\centering
\hspace{-3.5cm} \epsfxsize=12cm \epsfbox{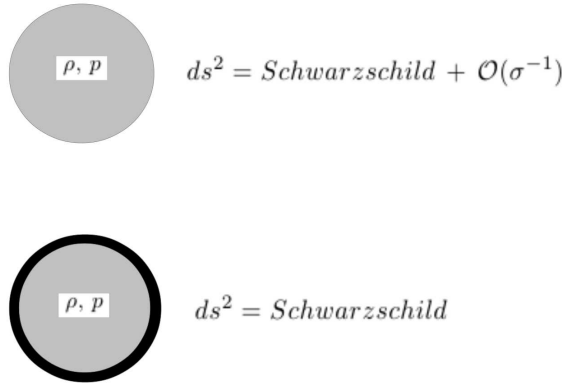}
\caption{Schematic picture of a brane-world star with and without a solid
thin outer crust: the interior is characterised by the density $\protect\rho 
$, pressure $p$ and star radius $R$. The exterior geometry by the ADM mass $%
\mathcal{M}$ and corrections proportional to $\protect\sigma ^{-1}$ when
there is no solid crust, and by the Schwarzschild geometry when there is a
solid crust which thickness $\Delta \sim {1}/{R\,\protect\sigma }$. }
\label{BWstar}
\end{figure}

The negative pressure shell has rather a tension and qualifies as a solid
(as illustrated in Fig.~\ref{BWstar}). The interpretation of a solid
material appearing as consequence of the extra dimension in the context of
brane-worlds was first advanced in Ref.~\cite{HomBrane}, which presented the
homogeneous counterpart of the Einstein brane~\cite{EinBrane}. Another
brane-world star, the perfect fluid material of which in the latest stages
of the collapses obeys the dark energy condition, was discussed in
Ref.~\cite{gergely2007}. In the present case however the region with negative
pressure is tiny and effectively acts as a solid crust, separating the inner fluid
from the vacuum exterior.
Moreover, in the crust all energy conditions hold, as they do everywhere inside
the star.

The thickness $\Delta $ of the solid crust is given by $\Delta^{-1}\sim {R\,\sigma }$,
showing thus that this ``solidification process'' in the outer crust due to
extra-dimensional effects becomes more important for compact stellar
distributions. Moreover, since the dimensionless parameter
$\Delta /\sigma^{-1/2}\sim \left( \sigma ^{1/2}R\right) ^{-1}\ll 1$ for astrophysical
stars, this crust has negligible thickness, falling below any physically
sensible length scale for astrophysical sources. We refrain from trying to
develop a detailed mechanism to realise the negative pressure, precisely
because the thickness falls below length scales for well-known physics.
In fact, the very existence of the crust could be masked by modifications of the
fundamental gravitational theory above the brane-world energy scale $\sigma $.
For testing such modifications, however precision measurements probing physics
above the scale set by $\sigma $ would be necessary.

As the stars with solid crust discussed in this paper are embedded in
vacuum, they do not radiate, strictly speaking.
In order to include radiation, one should in fact match the star interior with a
Vaidya exterior (containing radiation in the geometrical optics limit, or null dust).
Nevertheless, the emission of thermal radiation is allowed within our approximations,
similarly as our Sun (the exterior of which is also well approximated by a vacuum
Schwarzschild solution) can be well described like a black body emitting radiation
at approximately $5800\,$K.
One can then argue that a black body radiation outside our brane-world stars 
with solid crust should not affect the geometry significantly,
precisely like this radiation is negligible for the Sun in 4-dimensional GR.
Finally, the crust should be transparent to this thermal radiation, otherwise
it would accumulate energy and become quickly unstable.

At a technical level, the Weyl source functions $\mathcal{U}=\mathcal{U}(r)$
and $\mathcal{P}=\mathcal{P}(r)$ exhibit a discontinuity at the star surface 
$r=R$, where the radial effective pressure $\tilde{p}_{r}(R)=0$. Concerning
these discontinuities, it is known~\cite{germ} that
$\mathcal{U}(r)=\mathcal{P}(r)=0$ cannot hold everywhere on the brane if matter
is present inside a compact brane-world region (in our case, for $r<R$).
The question then arises whether a jump in $\mathcal{U}=\mathcal{U}(r)$ and
$\mathcal{P}=\mathcal{P}(r)$ at $r=R$ (see Figs.~\ref{anisoU} and~\ref{anisoP})
might signal some pathological behaviour of the extension into the bulk of the stellar
solution. We will argue, that this is not the case. Most brane-world models,
including the present one, assume the brane is a (Dirac $\delta $-like)
discontinuity along the extra dimension. The material sources on the brane
are regular, however, as the brane itself represents a hypersurface, they
could only be extended into the fifth dimension as Dirac $\delta $-like
distributions. The way to avoid that and have a regular stellar matter
distribution in all dimensions would be to consider a thick brane (like a
very narrow Gaussian of thickness, say, of order $\sigma ^{-1/2}$ along the
extra spatial dimension), on top of which the regular distribution of
brane-world matter could be placed (for more details, see Ref.~\cite{germaniC}).
The ``brane-world limit'' could then be obtained by assuming $\sigma ^{-1/2}$
is much shorter than any length scale associated with brane-world matter.
The approach followed in this paper instead, is based on the effective 4-dimensional
brane-world equations obtained from the $\delta $-like brane energy density in the bulk,
and a check that the behaviour of all physical variables is sufficiently
well-behaved. Specifically, one can see from the Figs.~\ref{anisoU}
and~\ref{anisoP} that the discontinuities in $\mathcal{U}=\mathcal{U}(r)$
and $\mathcal{P}=\mathcal{P}(r)$ at $r=R$ are negligibly small, most likely
generated by the $\delta $-like brane approach.

It is commonly believed that brane-world modifications to a star should be
detected through the long range behaviour, manifesting themselves in the
weak field regime. Orbital motions or gravitational lensing could then
provide information about the parameters of the stars, like its mass,
rotation, quadrupole moment and so on. Strong field effects, like those
occurring in the inner edge of an accretion disk, supposed to be at the
innermost stable circular orbit also depend only on the exterior geometry.
Stellar astrophysical processes leading to electromagnetic radiation could
also be slightly modified in brane-worlds (although standard model fields
remain 4-dimensional, gravity is changed, hence the equilibrium between
radiation pressure and gravitational attraction, for example, is shifted).
In fact such constraints were already derived for the tidally charged brane
black hole~\cite{dadhich}, and include constraints on the tidal charge from
the deflection of light~\cite{BHL, TidalLens, TidalLens2}, from the radius
of the first relativistic Einstein ring due to strong lensing~\cite{HZsfikut}
and from the emission properties of the accretion disks, including
the energy flux, the emission spectrum and accretion efficiency~\cite{accr}. 

The importance of the results presented in this paper relies in explicitly
illustrating that no matter how severe constraints from lensing or other
tests are derived for brane-world stars with exteriors depending on
brane-world parameters, the existence of the brane-world stars
cannot be ruled out, as their exterior could be the same as in GR. The
brane-world stars presented in this paper composed of a fluid with
physically reasonable properties and having a solid crust, immersed into a
vacuum Schwarzschild region on the brane, precisely exhibits this property
of indistinguishability.
%
%
%
%
%

\section*{Acknowledgements}

The research of L~\'{A}~G was supported by the European Union and State of
Hungary, co-financed by the European Social Fund in the framework of the T%
\'{A}MOP-4.2.4.A/2-11/1-2012-0001 \textquotedblright National Excellence
Program\textquotedblright .

\end{document}